\documentclass[12pt,notitlepage]{revtex4-1} % aps,groupedaddress   ---   prb,preprint   ---    12pt,notitlepage
\usepackage{amsmath}  % needed for \tfrac, \bmatrix, etc.
\usepackage{amsfonts} % needed for bold Greek, Fraktur, and blackboard bold
\usepackage{graphicx} % needed for figures

\usepackage[svgnames]{xcolor}
\usepackage{subcaption}

\newcommand{\uvec}[1]{\boldsymbol{\hat{\textbf{#1}}}}

\begin{document}

% Be sure to use the \title, \author, \affiliation, and \abstract macros
% to format your title page.  Don't use lower-level macros to  manually
% adjust the fonts and centering.

\title{Beyond the magnetic field of a finite wire: a teaching approach using the superposition principle}
% Extrapolating the magnetic field produced by a finite wire: a teaching resource to apply the Biot-Savart law exploiting the superposition principle
%Beyond the magnetic field produced by a finite wire: a teaching resource exploiting the superposition principle
% In a long title you can use \\ to force a line break at a certain location.

\author{J.E. Garc\'ia-Farieta}
\email{joegarciafa@unal.edu.co}
\altaffiliation[permanent address: ]{Carrera 45 No. 26-85, Bogot\'a, Colombia} % optional second address
% If there were a second author at the same address, we would put another 
% \author{} statement here.  Don't combine multiple authors in a single
% \author statement.
\affiliation{Departamento de F\'isica, Universidad Nacional de Colombia - Sede Bogot\'a, 11001}
% Please provide a full mailing address here.

\author{A. Hurtado-M\'arquez}
\email{ahurtado@udistrital.edu.co}
\affiliation{Facultad de Ciencias y Educaci\'on, Proyecto Curricular de Licenciatura en f\'isica - Grupo de Investigaci\'on FISINFOR, Universidad Distrital Francisco Jos\'e de Caldas, Bogot\'a, 11021-110231588}

{
\let\clearpage\relax
\maketitle
}

\section{Introduction}

The traditional lecture-based scheme of teaching physics have shown to be ineffective in several aspects \cite{Sokoloff,Mcdermott, Welzel1997, Hake, Good_2018, Li_PhysRevPhysEducRes, DancyPhysRevSTPER}. It results in a lack of motivation, with low impact on students, to address more complex physical systems than the ones presented in textbooks. A clear example in university-level education is the electromagnetics course, in particular when the magnetic field concept is introduced \cite{Tasoglu_2014, Herrmann_1991, Guisasola_2004}.\\

Empirical laws of magnetism are usually discussed after a phenomenological description of the magnetic interaction and its relation with electric currents as magnetic sources. The textbooks show in that section how to apply the Biot-Savart law to evaluate the magnetic field in any point in space due to a little element of a current-carrying wire of arbitrary shape. Then, it is exemplified considering a straight finite wire and evaluating the magnetic field $\mathbf{B}$ at a specific point, however, it involves several assumptions, which make the problem easy to solve, but which are not discussed in detail, such as: i) choosing a symmetry point to evaluate $\mathbf{B}$; ii) applying the superposition principle to simplify vector operations; and iii) reducing the number of dimensions of the system, from 3D to 2D or 1D. Even if it can be justified, it makes difficult for students to solve some of the exercises proposed at the end of the textbook section, that aim to explore the magnetic field of more complex systems like an infinite wire, square loop, polygons of $n$ sides, circular loop among others. Omitting these key points to evaluate the magnetic field in such systems, leads students to use incorrect assumptions, while they try to adapt the methodology previously used in the simple case of a finite wire, resulting in algebra tricks or unreasonable results.\\

In this article, the expression of the magnetic field for a finite wire is used in a `inductive' way to obtain the magnetic field created by several configurations. This procedure is easy generalized from a computational approach, to evaluate the field in any point of space by using the analytic expression of the finite wire $n$-times and exploiting the superposition principle in those configurations. This hybrid methodology allows students to improve their analytic skills by extending the result of simple cases, as long as the assumptions made have been discussed. It also provides a motivation to create their own models and use programming tools to visualize the patterns of the magnetic field lines in different configurations, that are not always possible to perform in a laboratory demonstration.

\begin{figure}[h!]
\centering
\includegraphics[scale=0.5]{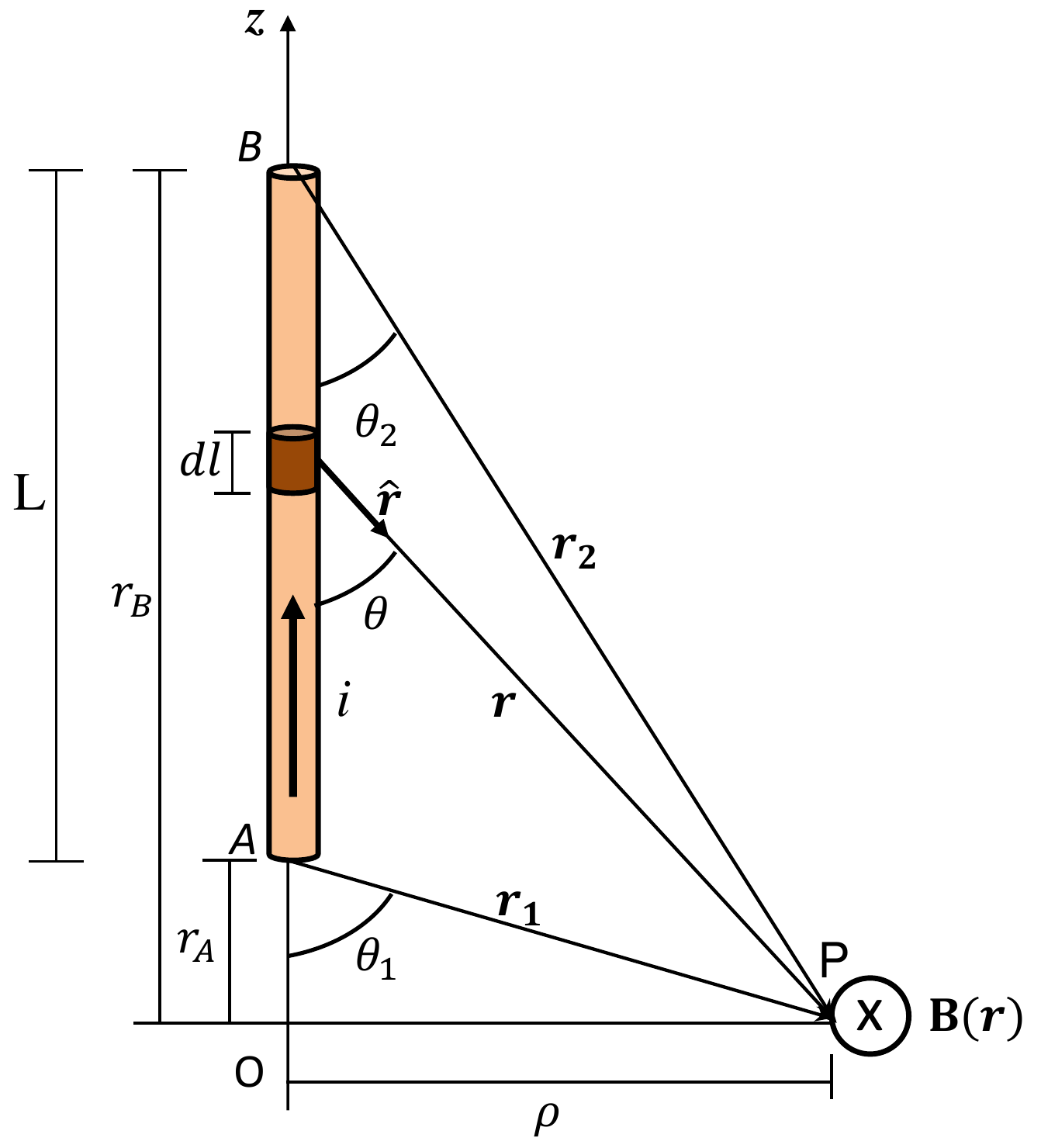}
\caption{Finite straight wire carrying a steady current along $z$-axis. The magnetic field is evaluated at the point $P$.}
\label{fig:system_finite_wire1}
\end{figure}

\section{A finite straight wire}
The Biot-Savart law describes the magnetic field $d\mathbf{B}$ at any point in space due to an element $d\mathbf{l}$ of a current-carrying wire as follows:
\begin{equation}\label{eq:Biot-Savart_Law}
    d\mathbf{B}=\frac{\mu_0I}{4\pi}\frac{d\mathbf{l}\times\uvec{r}}{r^2},
\end{equation}
where $\mu_0$ is the vacuum permeability and $\hat{\mathbf{r}}$ is an unit vector along $\mathbf{r}$. For a thin and straight wire carrying a current $I$ placed along the $z$-axis, as shown in Fig. \ref{fig:system_finite_wire1}, the magnetic field at an arbitrary point $P$ is given by:
\begin{equation}\label{eq:dB_FiniteWire}
d\mathbf{B}=\frac{\mu_0I}{4\pi}\frac{\rho dz}{\left(\rho^2+z^2\right)^{3/2}} \uvec{e}_\varphi,
\end{equation}
where cylindrical coordinates have been used for simplicity to map the three-dimensional space, taking advantage of the fact that $d\mathbf{l}$ and $\mathbf{r}$ are located in the same plain, causing that the magnetic field is always oriented along $\uvec{e}_\varphi$ for each point $P$, i.e, following concentric circle trajectories regardless of the segment size. By integrating Eq. \eqref{eq:dB_FiniteWire}, the total magnetic field produced by the wire can be expressed as:
\begin{equation}\label{eq:Integral_BiotSavart4}
\mathbf{B}=\frac{\mu_0I}{4\pi\rho}\left(\cos\theta_2-\cos\theta_1\right) \uvec{e}_\varphi\quad\text{where}\quad\cos\theta_i=\frac{z-z_i}{\sqrt{\rho^2+(z-z_i)^2}}\quad \text{and}\quad i=A,B.
\end{equation}
This equation gives a general description of the magnetic field for a current-carrying wire of finite length $L$, however, for the computational implementation it is useful to express it in Cartesian coordinates considering that the wire can be oriented in any direction. Using the law of cosines and vector decomposition, the magnitude of the magnetic field for an arbitrary located wire can be re-written as \cite{YuZhou}
\begin{equation}\label{eq:Integral_BiotSavart5}
B=\frac{\mu_0I}{4\pi}\frac{(r_1+r_2)\left(L^{2}-r_1^2-r_2^2+2r_1r_2\right)}{r_1 r_2 \sqrt{2r_1^2r_2^2+2r_1^2L^2+2r_2^2L^{2}-r_1^4-r_2^4-L^4}}.
\end{equation}
\begin{figure}[h!]
\centering
\includegraphics[scale=0.3]{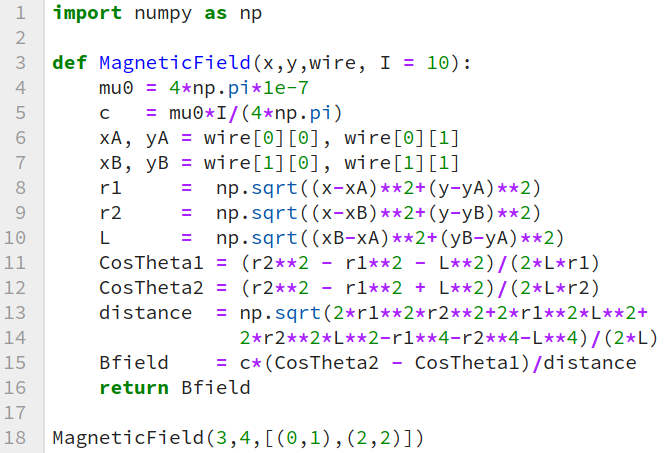}
\caption{Python routine to compute the magnitude of the magnetic field produced by a finite straight current-carrying wire arbitrarily oriented in space.}
\label{fig:code}
\end{figure}

Fig. \ref{fig:code} shows the function written in python language, whose inputs are: \emph{i)} the coordinates ($x,y$) where $B$ is evaluated, \emph{ii)} the position of the wire $(r_A,r_B)$, and \emph{iii)} the value of the electric current $I$. This short function allows students to explore how affects each one of the parameters in Eq. \eqref{eq:Integral_BiotSavart5} and to visualize different configurations, taking advantage that python language is very friendly and has a wide set of tools that can be used without a deep programming knowledge.\\

\begin{figure}
\centering
\begin{subfigure}[b]{.46\textwidth}
\includegraphics[width=1.2\textwidth]{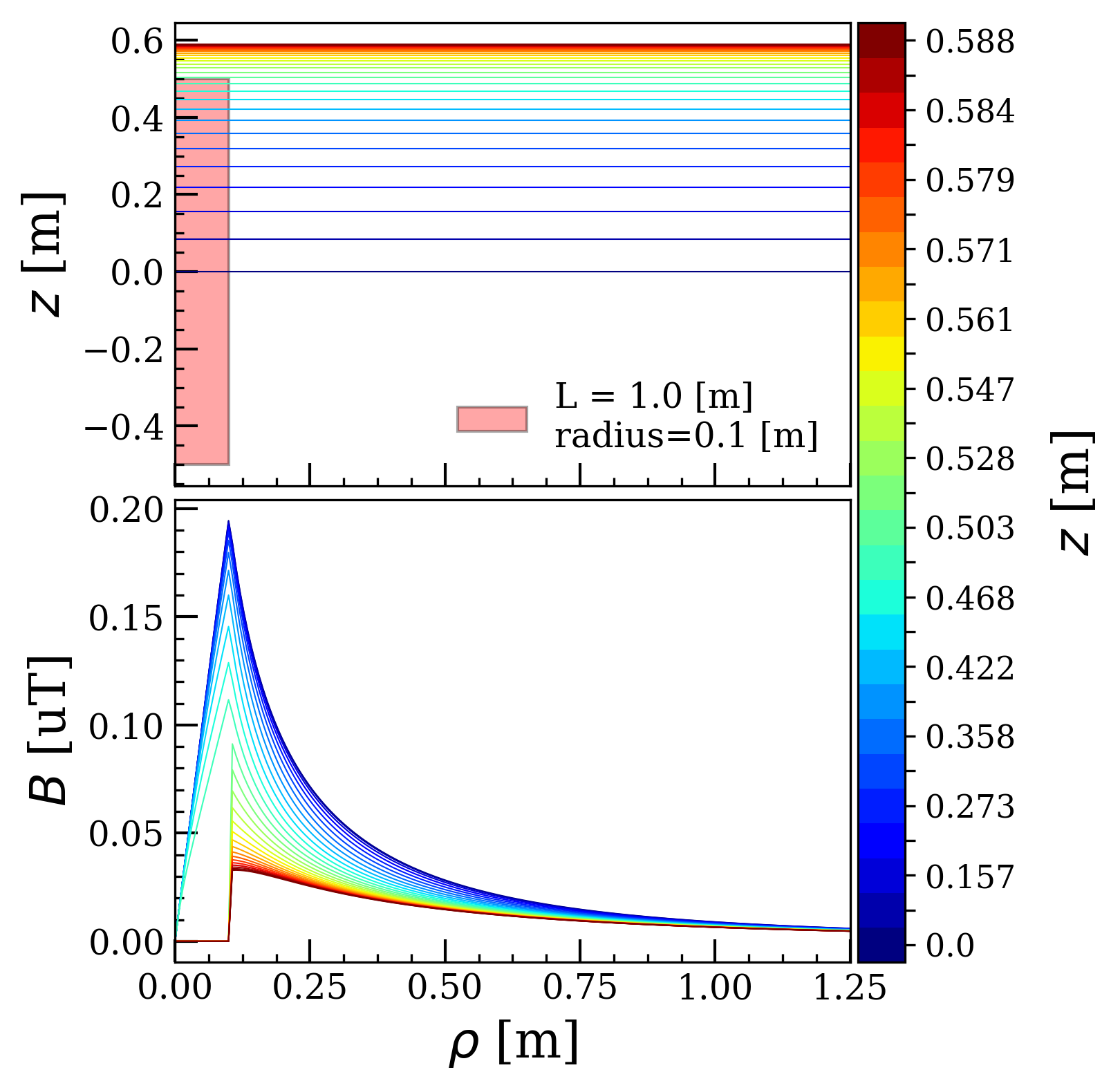}
\caption{}
\end{subfigure}\qquad
\begin{subfigure}[b]{.46\textwidth}
\includegraphics[width=0.8\textwidth]{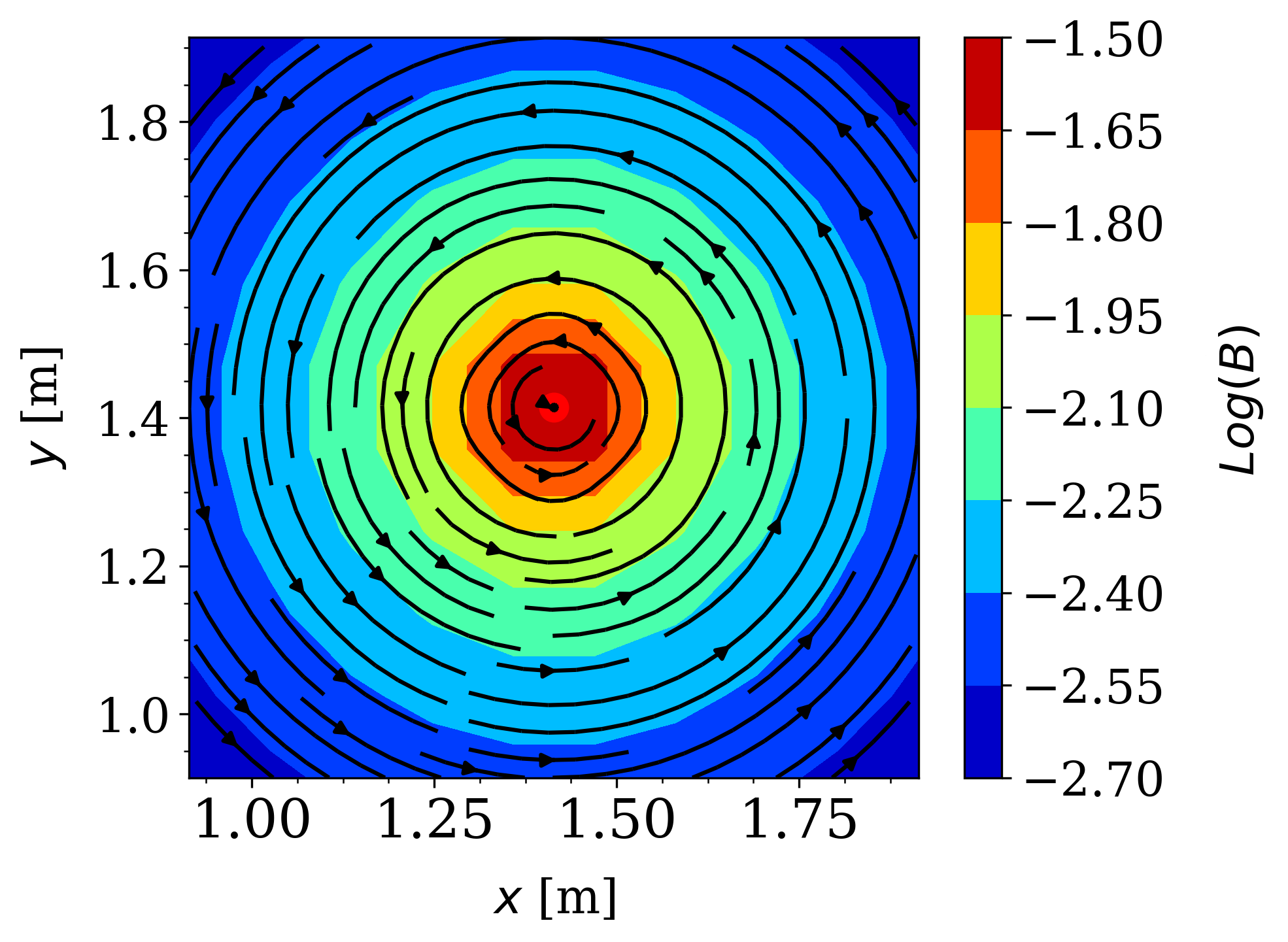}
\includegraphics[width=0.74\textwidth]{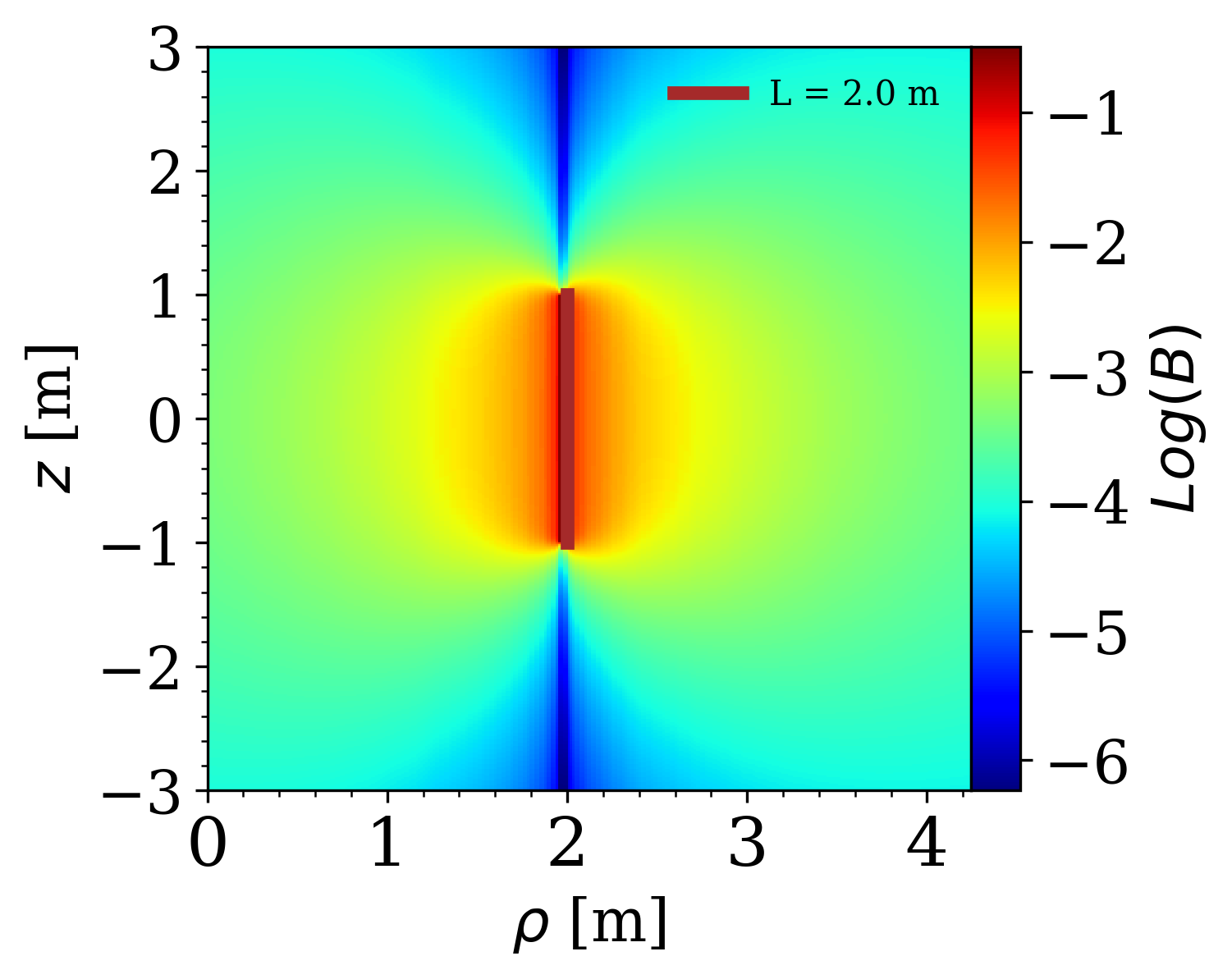}
\caption{}
\end{subfigure}
\caption{\emph{a)} magnitude of the magnetic field created by a current-carrying wire as a function of the distance $\rho$; \emph{b)} intensity map of the total magnetic field in a cross section view (\emph{upper}) and in a parallel plane to the wire (\emph{lower}).}
\label{fig:finite_wire}
\end{figure}

Fig. \ref{fig:finite_wire} shows the plot of evaluating $B$ in a meshgrid for a wire aligned along the $z$-axis. Subfigure \emph{a)} displays $B$ as a function of the distance $\rho$ for points at different heights as shown by the colorbar. Inside the wire, the magnitude of the field increases linearly with distance until reaching the wire radius, then outside the wire, it decreases inversely with the distance. The field at points within the radius of the wire but above/below it tends to zero as expected. Subfigure \emph{b)} shows the intensity map of the total magnetic field in a cross section view of the wire (\emph{upper}) and in the $z-\rho$ plane (\emph{lower}). The direction of the arrows indicate that the current is coming out from the plane following the right-hand rule.

%----------------------------------------------
\section{Application of the superposition principle}
\begin{figure}[h!]
\centering
\includegraphics[scale=1.0]{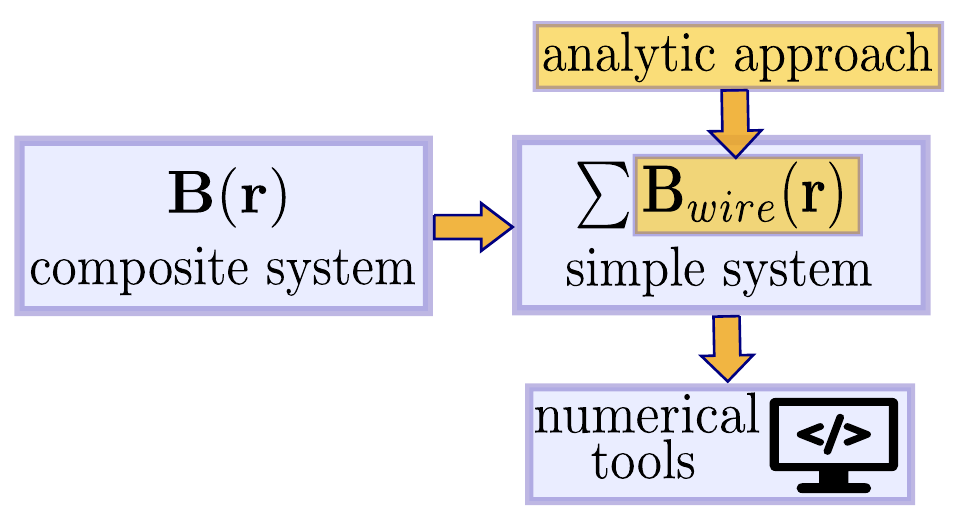}
\caption{Basic idea to obtain the magnetic field of a system using the superposition principle.}
\label{fig:diagram}
\end{figure}

The superposition principle states that the field created by different sources, e.g., by two or more currents, simply added together as vectors. Applying the superposition of magnetic fields before any calculations makes a transition to problem-solving easier. This approach is represented in Fig. \ref{fig:diagram}. Starting from the analytic expression for a finite wire, the magnetic field of a composite system can be expressed as the one created by an arrangement of $n$-wires, being simple to compute numerically by iterating Eq. \eqref{eq:Integral_BiotSavart5}, or equivalent the python routine (see Fig. \ref{fig:code}).\\

\begin{figure}[h!]
      \includegraphics[width=0.4\textwidth]{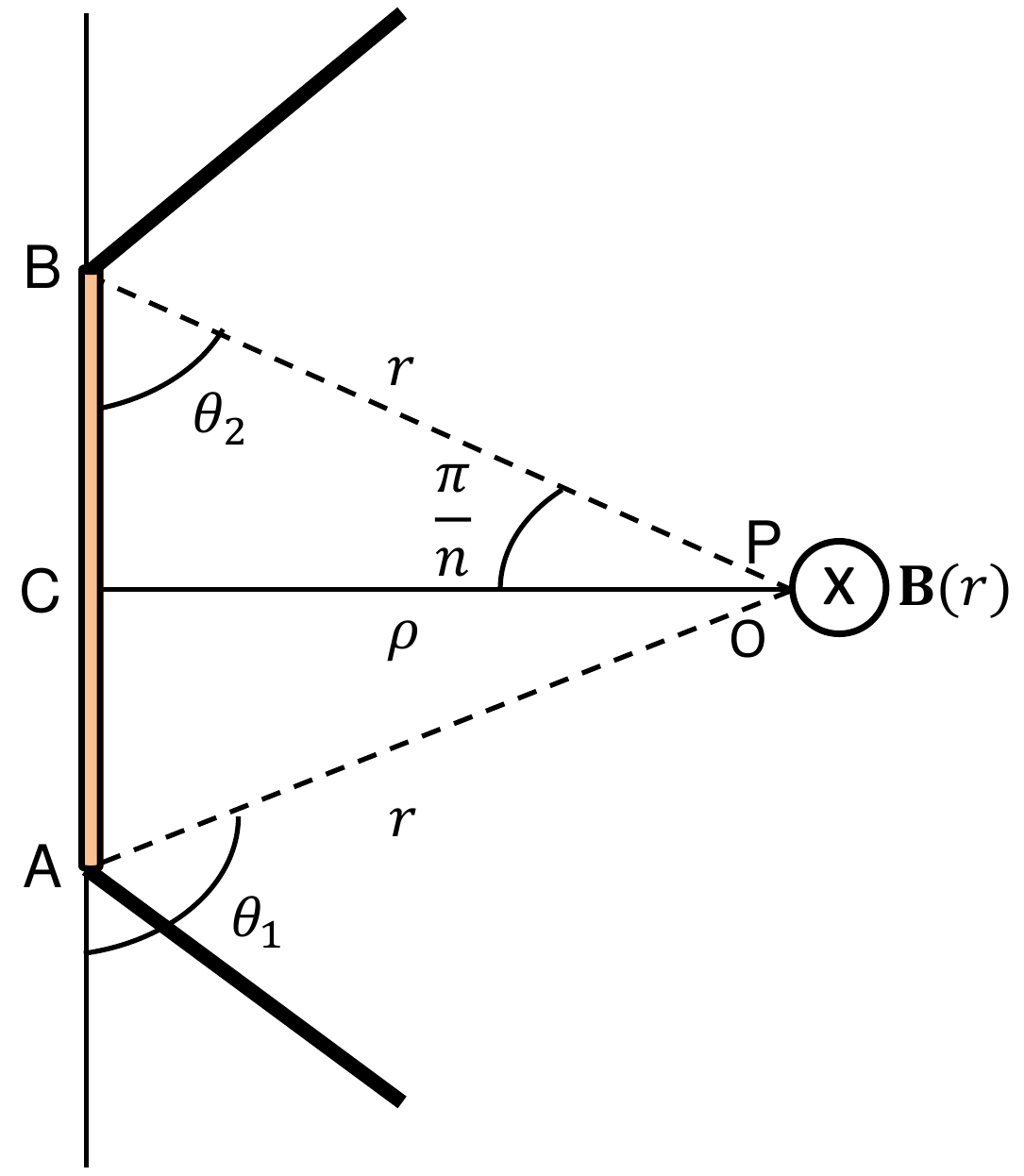}
  \caption{Systems considered for several configurations.}
  \label{fig:systems_wires}
\end{figure}

To illustrate this idea, let's consider a $n$-side regular polygon carrying a current $I$. Each side corresponds to a wire, then Eq. \eqref{eq:Integral_BiotSavart4} can be used to find $B$ at any point of space. By instance, the field at the center of the polygon can be obtained using Fig. \ref{fig:systems_wires}, where $AB$ is one of the sides and $\rho$ is the radius of the inscribed circle. The angles $\theta_2$ and $\theta_1$ are derived from the geometrical construction, taking into account that the inner angle of the polygon is given by $\angle AOB=2\pi/n$, thus $\angle BOC=\pi/n$. Considering the $n$ wires and using angle properties, the magnitude of the magnetic field is given by:
\begin{eqnarray}\label{eq:Bfield_polygon}
B&=&n\frac{\mu_0i}{4\pi \rho}\left[\cos\left(\frac{\pi}{2}-\frac{\pi}{n}\right)-\cos \left(\frac{\pi}{2}+\frac{\pi}{n}\right)\right],\nonumber\\
&=&n\frac{\mu_0i}{2\pi \rho}\sin\left(\frac{\pi}{n}\right).
\end{eqnarray}
As example, let's evaluate this expression in two particular cases. For a square loop of side $l$: $n=4$ and $\rho=l/2$, thus Eq. \eqref{eq:Bfield_polygon} becomes 
\begin{equation}\label{eq:square_loop}
B=\frac{2\sqrt{2} \mu_0i}{\pi l},
\end{equation}
whereas for a circular loop of radius $r$: $n\rightarrow\infty$ and $\rho=r$, then the magnetic field at its center is given by
\begin{eqnarray}
B&=&\lim_{n\rightarrow\infty} n\frac{\mu_0i}{2 \pi r} \sin\left(\frac{\pi}{n}\right),\nonumber\\
&=&\frac{\mu_0i}{2r}\lim_{n\rightarrow\infty} \frac{\sin\left(\frac{\pi}{n}\right)}{\left(\frac{\pi}{n}\right)}\nonumber\\
&=&\frac{\mu_0 i}{2 r}.
\end{eqnarray}

%============================================================
\section{Numerical results}
\begin{figure}[h!]
\centering
\includegraphics[width=0.6\textwidth]{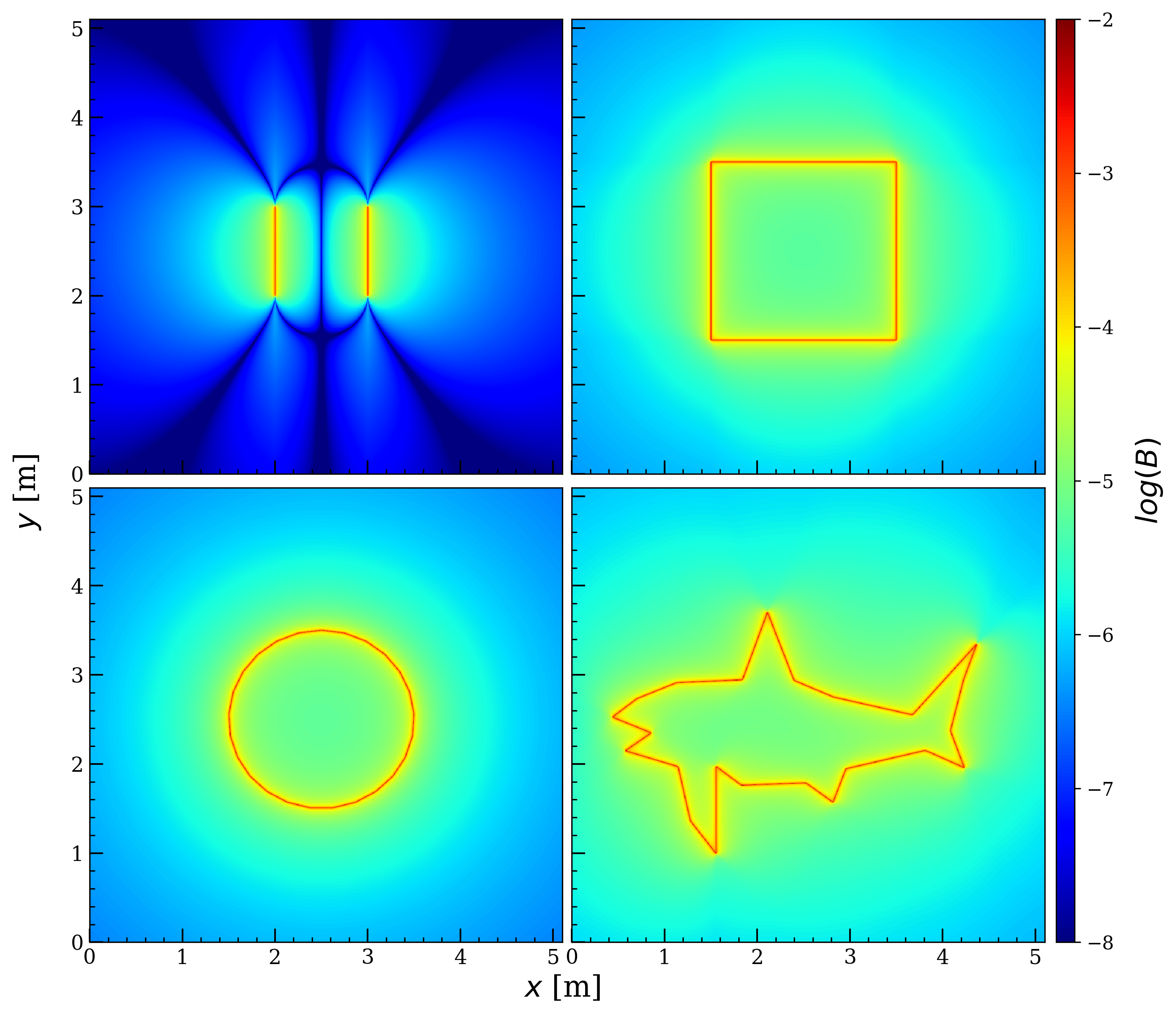}
\caption{Magnetic field distribution of different composite configurations evaluated with the superposition principle by using the expression of a finite wire.}
\label{fig:field_distributions_shapes}
\end{figure}

Using the python code for a wire, it is simple to extend the result to more elaborate geometric distributions. Thus the student can obtain the magnetic field at any point in space, visualize and characterize it, even without deriving the analytic expression for the system. Fig. \ref{fig:field_distributions_shapes} shows the results for four different configurations: a) two finite parallel wires; b) square loop, c) circular loop, and d) an irregular arrangement with a fish shape.\\

The systems can be described in terms of the field intensity and identifying if there are regions where the field is canceled. The students can also change the input parameters and explore its impact graphically, by example varying the electric current (value and direction), the wire lengths and also recover the configurations proposed in textbooks to compare with the analytic solution.\\

As an exercise, one can suggest to create their own models in more than two dimensions for specific distributions, and explore what happens changing the parameters. This inductive process leads to a interesting activity, since an arbitrary system is defined only in terms of the coordinates of the wires, thus an arraignment in 3D can show even more complex, and at the same time more realistic, configurations. For example, it can be used to find the magnetic field of solenoids, loops configurations like Helmholtz coils or even a toroidal system, before performing exhausting theoretical calculations.

\section{Conclusion}

We have presented a practical methodology to use the superposition principle of magnetic fields starting from a finite current-currying wire. A short python routine was designed to illustrated how to apply the superposition principle to obtain the magnetic field of complex configurations in any point in space, without need of an explicit analytic expression. It can be used as a complement of classroom experiments, motivating the discussion on magnetic phenomena and providing a better understanding of the Biot-Savart law. This approach allows students to explore systems with different level of complexity, from regular polygons to very irregular distributions as illustrated in Fig. \ref{fig:field_distributions_shapes}, combining basic analytic skills with computational tools.

\end{document}